%% file: eyetracking.tex
\documentclass[runningheads,a4paper]{llncs}
\usepackage{graphicx}
\usepackage{float}
\usepackage{multirow}
\usepackage{wrapfig}

\graphicspath{{img/}}

\newcommand{\keywords}[1]{\par\addvspace\baselineskip
\noindent\keywordname\enspace\ignorespaces#1}

\begin{document}
\mainmatter

\title{Investigating the Process of Process Modeling with Eye Movement
Analysis\thanks{The final publication is available at Springer via
http://dx.doi.org/10.1007/978-3-642-36285-9\_46}\\ \vspace{0.3cm} \small Full
Paper \vspace{-0.5cm} }
\titlerunning{Investigating the Process of Process Modeling with Eye Movement Analysis}
\authorrunning{Jakob Pinggera et al.}

% \author{Jakob Pinggera\inst{1} \and Marco Furtner\inst{1} \and Markus
% Martini\inst{1} \and Pierre Sachse\inst{1} \and Katharina Reiter\inst{1} \and
% Stefan Zugal\inst{1} \and Barbara Weber\inst{1}}
% 
% \institute{
% 	University of Innsbruck, Austria\\
% \email{firstname.lastname@uibk.ac.at, katharina.reiter@student.uibk.ac.at}}	
\author{Jakob Pinggera\inst{1} \and Marco Furtner\inst{2} \and Markus
Martini\inst{2} \and Pierre Sachse\inst{2} \and Katharina Reiter\inst{2} \and
Stefan Zugal\inst{1} \and Barbara Weber\inst{1}}

\institute{
	Department of Computer Science, University of Innsbruck, Austria\\
\email{jakob.pinggera|stefan.zugal|barbara.weber@uibk.ac.at}	
\and
	Department of Psychology, University of Innsbruck, Austria\\
\email{marco.furtner|markus.martini|pierre.sachse@uibk.ac.at,
katharina.reiter@student.uibk.ac.at} }	
\maketitle

\vspace{-0.7cm}
\begin{abstract}
Research on quality issues of business process models has recently begun to
explore the process of creating process models by analyzing the modeler's
interactions with the modeling environment. In this paper we aim to complement
previous insights on the modeler's modeling behavior with data gathered by
tracking the modeler's eye movements when engaged in the act of modeling. We
present preliminary results and outline  directions for future research to
triangulate toward a more comprehensive understanding of the process of process
modeling. We believe that combining different views on the process of process
modeling constitutes another building block in understanding this process that
will ultimately enable us to support modelers in creating better process models.

\vspace{-0.3cm}
\end{abstract}

\input{introduction}

\input{background}
\input{methods}
\input{datacollection}
\input{discussion}

\input{relatedWork}

\input{summary}

\bibliography{literature}
\bibliographystyle{splncs}
\end{document}

%% file: introduction.tex
\vspace{-0.9cm}
\section{Introduction}\label{sec:intro}
Considering the heavy usage of business process modeling in all types of business
contexts, it is important to acknowledge both the relevance of process models and
their associated quality issues. On the one hand, it has been shown that a good
understanding of a process model has a positive impact on the success of a
modeling initiative~\cite{kock2009communication}. On the other hand, actual
process models display a wide range of problems that impede their
understandability~\cite{mendling-lnbip}. Clearly, an in-depth understanding of
factors influencing process model quality is in demand.

Most research in this area puts a strong emphasis on the \textit{product} of the
process modeling act, i.e., the process model, (e.g.,~\cite{van2000verification}). Other
works---instead of dealing with the quality of individual models---focus on the
characteristics of modeling languages (e.g.,~\cite{DBLP:journals/tse/Moody09}).
Recently, research has begun to explore another dimension presumably affecting
the quality of business process models by looking into the \textit{process of creating a
process model} (e.g.,~\cite{Sof+11,PZW+11,PSZ+12,CVR+12}). Thereby, the
focus has been put on the \textit{formalization phase}, in which a process
modeler is facing the challenge of constructing a syntactically correct model reflecting
a given domain description~\cite{hoppenbrouwers05er}. Our research can be
attributed to the latter stream of research.

This paper contributes to our understanding of the process of process modeling
(PPM) by combining modeling phase diagrams~\cite{PZW+11} with data collected by
analysing the modeler's eye movements. We demonstrate the feasibility of using
eye movement analysis to complement existing analysis techniques for the PPM by
presenting preliminary results and outline directions for future work. We
postulate that by analysing the PPM from different viewpoints, a more
comprehensive understanding of the process underlying the creation of process
models can be obtained, facilitating the creation of modeling environments that
support modelers in creating high quality models. Similarly, improved knowledge
about the PPM can be exploited for teaching students in the craft of modeling.

The paper is structured as follows. Section~\ref{sec:backgrounds} presents
backgrounds on the PPM. Section~\ref{sec:methodology} introduces eye movement
analysis. Section~\ref{sec:datacollection} describes the conducted modeling
sessions, whereas Section~\ref{sec:example} presents preliminary results. The
paper is concluded with related work in Section~\ref{sec:relatedwork} and a
summary in Section~\ref{sec:summary}.

%% file: background.tex
\vspace{-0.3cm}
\section{Background}\label{sec:backgrounds}
\vspace{-0.2cm}
This section describes backgrounds of the PPM and illustrates how the PPM can be
visualized using modeling phase diagrams. 
\vspace{-0.3cm}
\subsection{The Process of Process Modeling} 
During the formalization phase process modelers are creating a formal process
model reflecting a given textual domain description by interacting with the process
modeling environment~\cite{hoppenbrouwers05er}. At an
operational level, the modeler's interactions with the tool would typically
consist of a cycle of the three successive phases of (1) comprehension (i.e.,\
the modeler forms a mental model of domain behavior), (2) modeling (i.e.,\ the
modeler maps the mental model to modeling constructs), and (3) reconciliation
(i.e.,\ the modeler reorganizes the process model)~\cite{Sof+11,PZW+11}.
 
\noindent\textbf{Comprehension.} According to~\cite{NeSi72}, when facing
a task, the problem solver first formulates a mental representation of the
problem, and then uses it for reasoning about the solution and which methods
to apply for solving the problem. In process modeling, the task is to create
a model which represents the behavior of a domain. The process of forming
mental models and applying methods for achieving the task is not done in one
step applied to the entire problem. Rather, due to the limited capacity of
working memory, the problem is broken down to pieces that are addressed
sequentially, chunk by chunk~\cite{Sof+11,PZW+11}.

\noindent\textbf{Modeling.} The modeler uses the problem and solution developed
in working memory during the previous comprehension phase to materialize the
solution in a process model (by creating or changing it)~\cite{Sof+11,PZW+11}.
The modeler's utilization of working memory influences the number of modeling
steps executed during the modeling phase before forcing the modeler to revisit
the textual description for acquiring more information~\cite{PZW+11}.

\noindent\textbf{Reconciliation.} After modeling, modelers typically
reorganize the process model (e.g.,\ renaming of activities) and utilize the
process model's \emph{secondary notation} (e.g.,\ notation of layout,
typographic cues) to enhance the process model's
understandability~\cite{Petr95}. However, the number of reconciliation
phases in the PPM is influenced by a modeler's ability of placing elements
correctly when creating them, alleviating the need for additional
layouting~\cite{PZW+11}.

\vspace{-0.4cm} 

\subsubsection{Modeling Phase Diagrams (MPD).}
In order to facilitate the systematic investigation of the PPM, Cheetah
Experimental Platform (CEP) has been developed~\cite{PiZW10}. In particular, a
basic process modeling editor is instrumented to record each user's
interactions in an event log, describing the creation of the process model step
by step. When modeling in a process modeling environment, process modeling
consists of adding nodes and edges to the process model, naming or renaming
activities, and adding conditions to edges. In addition to these interactions, a
modeler can influence the process model's secondary notation, e.g.,\ by laying
out the process model using move operations for nodes or by utilizing bendpoints
to influence the routing of edges, see~\cite{PZW+11} for details. By capturing
all of the described interactions with the modeling tool, we are able to
\emph{replay} a recorded modeling process at any point in time\footnote{A replay
demo is available at http://cheetahplatform.org}~\cite{PZW+11}.

\begin{figure}[t]
\begin{center}
  \includegraphics[width=0.9\textwidth]{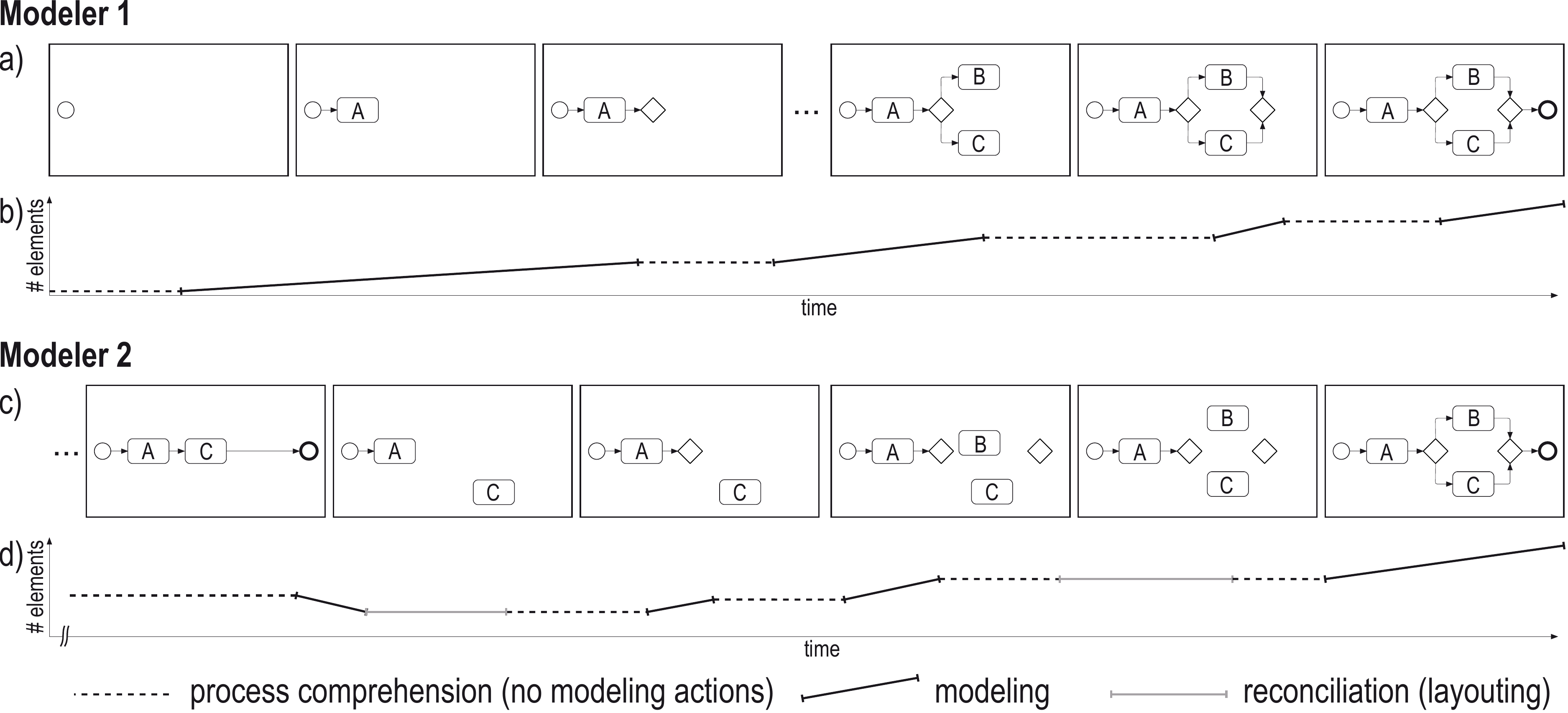}
  \vspace{-0.2cm}
  \caption{Two different PPM instances creating the same process
  model~\cite{PZW+11}}
  \label{fig:ppmoverview}
\end{center}
\vspace{-0.4cm}
\end{figure}

In~\cite{PZW+11} a technique for visualizing the PPM is proposed by mapping the modeler's interactions with the
modeling environment to the phases described above.
Fig.~\ref{fig:ppmoverview}a shows several states of a typical modeling process as
it can be observed during replay. Fig.~\ref{fig:ppmoverview}c shows the states of
a different modeling process that nonetheless results in the \emph{same} model.

To obtain a better understanding of the modeling process and its phases, we
supplement model replay with a \emph{modeling phase diagram}, quantitatively
highlighting the three phases of modeling, comprehension, and reconciliation. It
depicts how the size of the model (vertical axis) evolves over time (horizontal
axis), as can be seen in Fig.~\ref{fig:ppmoverview}b and
Fig.~\ref{fig:ppmoverview}d for the modeling processes in
Fig.~\ref{fig:ppmoverview}a and Fig.~\ref{fig:ppmoverview}c, respectively. A
\emph{modeling phase} consists of a sequence of interactions to create or delete
model elements such as activities or edges. A modeler usually does not create a
model in a continuous sequence of interactions, but rather pauses after several
interactions to inspect the intermediate result and to plan the next steps.
Syntactically, this manifests in reduced modeling activity or even inactivity,
i.e, a \emph{comprehension phase}. Besides, modelers need to \emph{reorganize}
the model. Reconciliation interactions manifest in moving or renaming model
elements to prepare the next modeling interactions or to support their
comprehension of the model. A sequence of such interactions is a
\emph{reconciliation phase}.

% So, we can read from Fig.~\ref{fig:ppmoverview}b that the modeler created the
% model in a straight-forward series of modeling steps interrupted by periods of
% comprehension. The modeling process in Fig.~\ref{fig:ppmoverview}d shows a
% different approach. After some modeling, the modeler removes parts of the created
% model and moves an activity to make space for some control-flow constructs, as
% indicated by the reconciliation phase. Then, several model elements are placed
% and laid out before the model is completed. Note that the resulting models are
% identical. Yet, the phase diagrams show significant differences between both
% modeling processes. This illustrates the value of analyzing the modeling process
% in the described manner beyond the inspection of the process models themselves.

%% file: methods.tex
\vspace{-0.4cm}
\section{Eye Movement Analysis}\label{sec:methodology}
\vspace{-0.2cm}
Even though MPDs provide valuable insights into the PPM, the modeler's
cognitive processes are left in the realm of speculation. More specifically, in
a MPD the various phases are detected by classifying the modeler's interactions with the
modeling environment and aggregating them to the various phases of the
PPM~\cite{PZW+11}. Comprehension phases in a MPD are assessed by measuring the
duration not interacting with the modeling tool~\cite{PZW+11}. Thresholds
are utilized for differentiating between an actual comprehension phase and the
usual inactivity between creating model elements, i.e., the time it takes the
modeler to select a different tool and create the next model
element~\cite{PZW+11}. This draws a rather coarse grained picture of the PPM,
i.e., shorter comprehension phases are not detected. Similarly, the authors
in~\cite{PZW+11} claim that there are diverse reasons for comprehension phases.
On the one hand, the modeler might create an internal representation of the
modeling task presented as an informal description. On the other hand, the
modeler might be understanding the process model or inspecting it for
potential errors. In order to develop a more fine grained understanding of the
PPM, we propose the combination of different views on the PPM. Subsequently, we
introduce eye movement analysis, which is combined with the corresponding MPD to
triangulate toward a more comprehensive understanding of the PPM.

\vspace{-0.3cm}

\subsubsection{Eye Movements.}When creating a formal process model from an
informal specification, a modeler relies on his visual perception for reading the
task description and creating the process model using the modeling environment.
In this context, high-resolution visual information input is of special interest,
which is necessary for reading a word or seeing an element of the process model.
High-resolution visual information input can only occur during so-called
\textit{fixations}, i.e., the modeler fixates the area of interest on the screen
with the fovea, the central point of highest visual acuity~\cite{Posn95}.
Fixations can be detected when the velocity of eye movements is below a certain
threshold for a pre-defined duration~\cite{JaKa03}. Using eye fixations, we can
identify areas on the screen the modeler is focusing attention on~\cite{FuSa08},
e.g., the task description, features of the modeling environment or modeling
constructs.

In order to perform a detailed analysis, the modeler's eye movements need to be
quantified. For this purpose, several different parameters exist~\cite{Rau+12}.
In this study we focus on two of the most widely used eye movement
parameters~\cite{JaKa03}.

\noindent\textbf{Number of Fixations.} The number of fixations is calculated by
counting the number of fixations in a pre-specified timeframe on a certain area
on the screen. This allows researchers to compare the number of fixations
on certain areas on the computer screen, e.g., the task description versus the
process model.

\noindent\textbf{Mean Fixation Duration.} The mean duration of fixations is
calculated by measuring the durations of fixations on a certain area on the
screen in a predefined timeframe and calculating the average duration. Longer
durations could be interpreted toward deeper processing of
information~\cite{Rau+12}, but might indicate inactivity of the
participant if fixation durations become too long compared to the
participants usual fixation durations~\cite{Wol+11}.

%% file: datacollection.tex
\vspace{-0.4cm}
\section{Data Collection}\label{sec:datacollection}
\vspace{-0.2cm}
In order to test the feasibility of combining eye movement analysis with existing
research on the PPM, i.e., MPD, we designed modeling sessions with students of
computer science and information systems. Participants were recorded using an eye
tracker when translating an informal description into formal process model. 
\vspace{-0.7cm}

\subsection{Definition and Planning}
This section describes the definition and planning of the modeling sessions.

\noindent\textbf{Subjects.}
The targeted subjects should be familiar with business process management and
imperative process modeling notations. More specifically, they should have 
prior experience in creating process models using BPMN. We are not targeting
modelers who are not familiar with BPMN at all to avoid measuring their learning
instead of the modeling behavior.

\noindent\textbf{Objects.}
The modeling session was designed to collect PPM instances of students creating a
formal process model in BPMN from an informal description. The informal
description was formulated in German since all participants were native German
speakers, avoiding potential translation problems. The object that was to be
modeled is a process describing the handling of mortgage request by a
bank\footnote{Material download:
http://pinggera.info/experiment/EyeMovementAnalysis}. The process model consists
of 19 activities and contains the basic control flow pattern: sequence, parallel
split, synchronization, exclusive choice, simple merge and structured
loop~\cite{Aals+03}.

\noindent\textbf{Response Variables.}
As already mentioned in the previous section, we recorded the number of
fixations and the duration of fixations. The PPM instances were cut into several
parts as detailed in Section~\ref{sec:example} and subsequently analysed. CEP
recorded the PPM instances on an operational level permitting the generation of
a MPD for each PPM instance.

\begin{figure}[t] 
\begin{center}
  \includegraphics[width=0.7\textwidth]{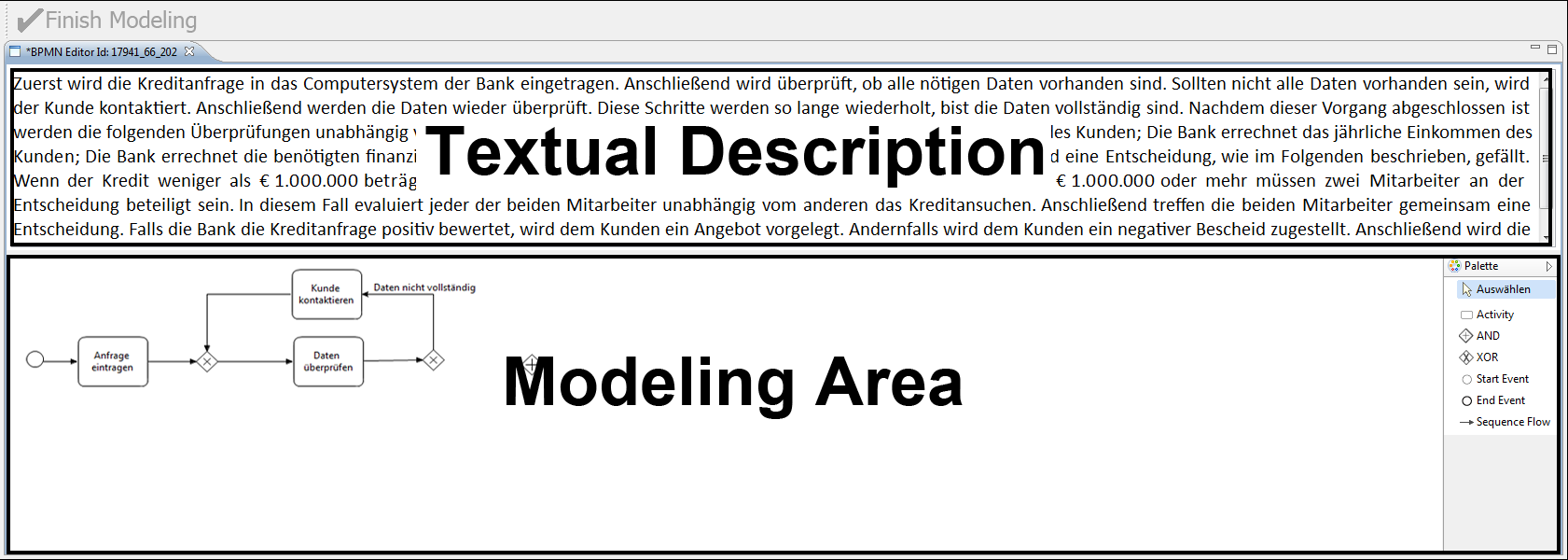}
  \caption{The BPMN Modeling Editor}
  \label{fig:cep}
\end{center}
\vspace{-0.4cm}
\end{figure}

\noindent\textbf{Instrumentation and Data Collection.}
CEP was utilized for recording the participants' PPM instances. To
mitigate the risk that PPM instances were impacted by complicated tools or
notations~\cite{Cra+00}, we decided to use a subset of BPMN. In order to
investigate the modeler's eye movements in the process model, but also in the
textual description we juxtapose the task description with the modeling area (cf.
\figurename~\ref{fig:cep}). Several pre-tests were conducted to ensure the
usability of the tool and the understandability of the task description. 

For performing the eye movement analysis we utilized a table mounted eye tracker,
i.e., Eyegaze Analysis System\footnote{http://www.eyegaze.com}, recording eye
movements using two binocular cameras positioned beneath a 17'' computer display 
with a frequency of 60 Hz each. Data recording is carried out with
the pupil center corneal reflection method~\cite{Ohn+02}. Data collection and
analysis is performed using NYAN 2.0\footnote{http://www.interactive-minds.com}.
The eye tracker is calibrated for each participant individually; calibrations are
accepted if the fixation accuracy shows an average drifting error of at most 0.25
inches. Two observation monitors allow watching both eyes separately while in the
process of eye-tracking to correct the sitting posture of participants to
recalibrate during recording if necessary.
\vspace{-0.2cm}

\subsection{Performing the Modeling Session}
This section describes the modeling sessions' execution.

\noindent\textbf{Experimental Execution.}
Since we have only access to a single eye tracker, each modeler has to be
recorded individually. 25 students of computer science or information systems participated
in the study. Each participant has taken classes on business process management
including the creation of business process models in BPMN. Modeling sessions were
conducted between February 2012 and May 2012 at the University of Innsbruck. The
experiment was guided by CEP's experimental workflow engine~\cite{PiZW10},
leading students through the modeling task, a concluding questionnaire and a
feedback questionnaire. Participation was voluntary; data collection was
performed anonymously.

\noindent\textbf{Data Validation.} Similar to~\cite{Men+07} we screened the
subjects for familiarity with BPMN by asking them whether they would consider
them to be very familiar with BPMN, using a Likert scale with values ranging from
\textit{Strongly disagree} (1) over \textit{Neutral} (4) to \textit{Strongly
agree} (7). The computed mean for familiarity with BPMN is 4.84 (slightly below
\textit{Somewhat Agree}). For confidence in understanding BPMN models, a mean
value of 5.76 was reached (slightly below \textit{Agree}). Finally, for perceived
competence in creating BPMN models, a mean value of 5.4 (between \textit{Somewhat
Agree} and \textit{Agree}) could be computed. Since all values range above
average, we conclude that the participating subjects fit the targeted profile.

%% file: discussion.tex
\vspace{-0.2cm}
\section{Combining MPD and Eye Movement Analysis}
\vspace{-0.2cm}
\label{sec:example} 
In this section we demonstrate the feasibility of combining eye movement analysis
with existing research on the PPM. Based on the data analysis
procedure described in Section~\ref{sec:dataAnalysis} two PPM instances
are presented and briefly discussed in Section~\ref{sec:mpds}. Preliminary results
from combining eye movement analysis with the corresponding MPD are discussed in
Section~\ref{sec:results}. 

\vspace{-0.3cm}
\subsection{Data Analysis}\label{sec:dataAnalysis}
\vspace{-0.2cm}
In this preliminary study, our focus was put on evaluating the feasibility of
combining eye movement analysis with existing research on the PPM,
e.g.,~\cite{PZW+11,PSZ+12,CVR+12}, and to investigate potential benefits of such
a combined analysis. For this purpose, we select two PPM instances for further
analysis. Similar to~\cite{PZW+11}, we use CEP to generate the MPD for each
modeler. In combination with CEP's replay feature we are able to gain an inital
understanding of the modeler's behavior. In order to validate and extend our
insights, we perform the eye movement analysis of the PPM. Since there are
several interesting timeframes exhibiting different characteristics in the PPM,
we manually separate the PPM into several, so-called, timeframes of interest
(TOI). TOIs are identified based on changes in the modeling behavior of the
participant, e.g., the modeler switches from adding model elements to resolving
problems. Please note that TOIs are identified for each modeler individually and
cannot be compared to TOIs of other modelers.

For each TOI in the PPM we distinguish between fixations on the textual
description and fixations on the modeling area (cf. \figurename~\ref{fig:cep}).
The relationship between fixations on the textual description and fixations on
the process model is expressed by calculating the percentage of fixations on the
textual description out of the total number of fixation (textual description and
process model). Additionally, we calculate the mean fixation duration for
fixations on the textual description and the mean fixation duration for fixations
on the process model.

\vspace{-0.3cm}

\subsection{PPM Examples}\label{sec:mpds}
\vspace{-0.2cm}
In this section we present the MPDs selected for further analysis.

\begin{wrapfigure}{r}{0.5\textwidth}
\vspace{-15pt}
\begin{center}
  \includegraphics[width=0.5\textwidth]{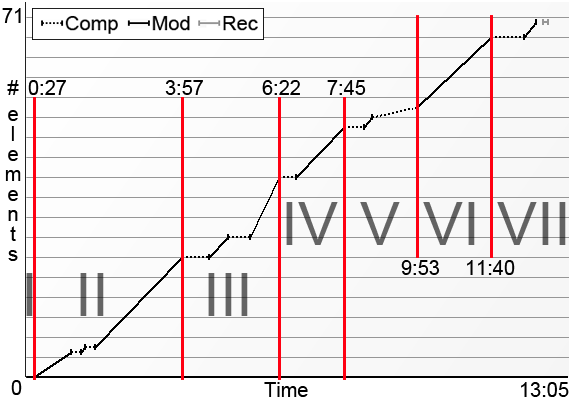}
  \caption{MPD of M1 with 7 TOIs}
  \label{fig:MPD_108}
\end{center}
\vspace{-15pt}
\end{wrapfigure}

\noindent\textbf{Modeler M1.} \figurename~\ref{fig:MPD_108} illustrates the PPM
of M1. In general, M1 produces the process model in a straight forward manner,
presumably with a clear conception of the resulting process model in mind. The
MPD shows several iterations of comprehension phases followed by long modeling
phases. TOI V constitutes an exception in the rather straight forward modeling
approach since an error is introduced, i.e., the modeler forgets about an
activity. The modeler immediately detects the problem and resolves it. The MPD
shows two comprehension phases which are only briefly interrupted by a modeling
phase\footnote{The number of elements in the process model can also change during
a comprehension phase in a MPD since several comprehension phases can be merged
when interrupted by brief modeling actions~\cite{PZW+11}.}. The PPM is
concluded by a brief reconciliation phase (the only one in this PPM).

\begin{table}[t]\scriptsize\sf
\centering
\begin{tabular}{l|r|r|r|r|r}
\hline
 & \multicolumn{2}{c|}{\textbf{Textual
Description}} & \multicolumn{2}{c|}{\textbf{Process Model}} & \multirow
{2}{4em}{\textbf{Fix. on Text[\%]}} \\ \textbf{TOI} & \textbf{Nr. of Fix.}
& \textbf{Mean Dur.[ms]} & \textbf{Nr. of Fix.} & \textbf{Mean Dur.[ms]} \\
\hline
I & 12 & 154 & 15 & 387 & 44.4\% \\
II & 174 & 164 & 442 & 199 & 28.2\%\\
III & 306 & 179 &  386 & 205 & 44.2\%\\
IV & 147 & 169 & 154 & 220 & 48.8\%\\
V & 228 & 204 & 215 & 237 & 51.5\%\\
VI & 19 & 156 & 415 & 192 & 4.4\%\\
VII & 31 & 188 & 194 & 249 & 13.8\%\\
\end{tabular}
\small
\caption{Eye Movement Analysis of M1}
\label{tab:modeler1}
\vspace{-0.2cm}
\end{table}

\begin{figure}[t]
\vspace{-0.2cm}
\begin{center}
  \includegraphics[width=\textwidth]{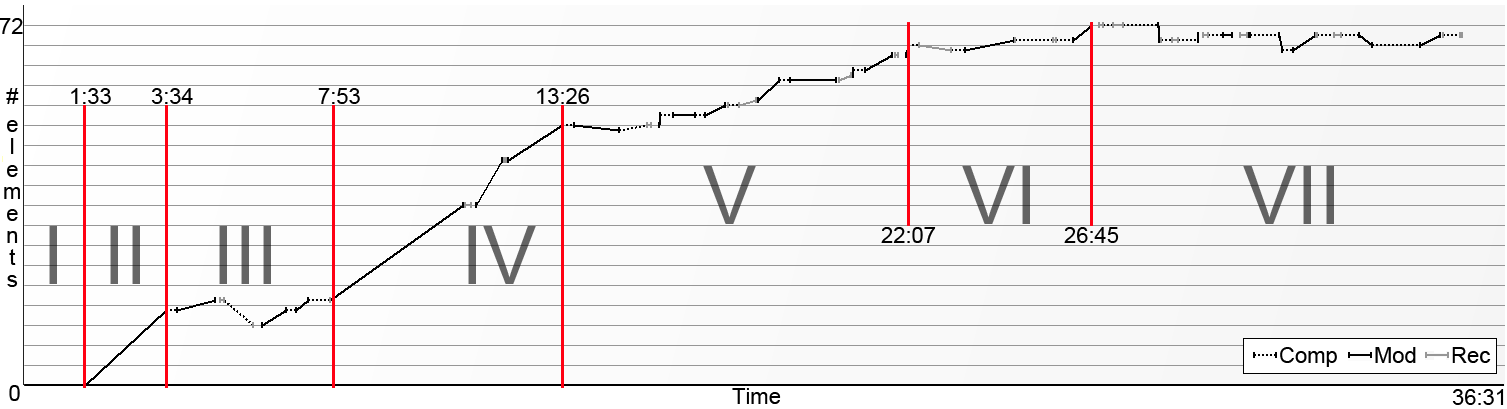}
  \caption{MPD of M2 with 7 TOIs}
  \label{fig:MPD_128}
\end{center}
\vspace{-0.2cm}
\end{figure}

\noindent\textbf{Modeler M2.} In contrast to M1, the MPD of M2 shows a very long
PPM (cf. \figurename~\ref{fig:MPD_128}). After a fast start, M2 experiences first
difficulties in TOI III, where M2 seems to struggle with introducing a loop in
the process model. After resolving this issues, the modeler returns to a fast
modeling style before experiencing problems toward the end of the PPM. In TOI V,
M2 adds parts of the process model just to remove them immediately on a trial and
error basis. This behavior changes in TOI VI when several long comprehension
phases and less delete operations can be observed. After achieving a complete
model at the end of TOI VI M2 checks the model for inconsistencies in TOI VII to
make occasional improvements.

\begin{table}[t]\scriptsize\sf
\centering
\begin{tabular}{l|r|r|r|r|r}
\hline
  & \multicolumn{2}{c|}{\textbf{Textual
Description}} & \multicolumn{2}{c|}{\textbf{Process Model}} & \multirow
{2}{4em}{\textbf{Fix. on Text[\%]}} \\ \textbf{TOI} & \textbf{Nr. of Fix.}
& \textbf{Mean Dur.[ms]} & \textbf{Nr. of Fix.} & \textbf{Mean Dur.[ms]} \\
\hline
I & 297 & 203 & 54 & 223 & 84.6\% \\
II & 88 & 165 & 368 & 204 & 19.3\%\\
III & 79 & 162 &  882 & 221 & 8.2\%\\
IV & 430 & 158 & 947 & 205 & 31.2\%\\
V & 491 & 173 & 1,710 & 209 & 22.3\%\\
VI & 95 & 177 & 921 & 233 & 9.4\%\\
VII & 319 & 149 & 1,760 & 211 & 15.3\%\\
\end{tabular}
\small
\caption{Eye Movement Analysis of M2}
\label{tab:modeler2}
\vspace{-0.3cm}
\end{table}

\vspace{-0.4cm}

\subsection{Preliminary Results}\label{sec:results}
\vspace{-0.2cm}
\tablename~\ref{tab:modeler1} and \tablename~\ref{tab:modeler2} show the
various eye movement parameters for each TOI of M1 and M2 respectively.
Subsequently, we present preliminary results deduced from combining
eye movement analysis with the corresponding MPD.

\noindent\textbf{Shorter Fixations when Reading.} When comparing mean fixation
durations it can be observed that mean durations are lower for fixations on the
task description compared to fixations on the process model. This finding is
consistent with results reported in literature indicating shorter fixations when
reading~\cite{Fur+11}.

\noindent\textbf{Fast and Focused Modeling.} In our previous research we observed
phases in the PPM when modelers created large chunks of their process models in
relatively short periods of time. We had the impression that modelers had a clear
picture of the PPM in mind, often alleviating them from subsequent reconciliation
phases since they placed model elements at strategic places right from the
beginning~\cite{PZW+11}. In a MPD, those periods are mostly indicated by long and
steep modeling phases. Also M1 and M2 exhibit such phases. M1 starts with long
modeling phases in TOI II and has another long modeling phase in TOI VI. M2 has
similar phases in TOI II and TOI IV. All TOIs have low mean fixation durations,
indicating a lower perceived complexity of the situation at hand compared to
other TOIs~\cite{UnRo90}. The lower perceived complexity, in turn, allows
modelers to consider additional model characteristics like the process model's
secondary notation right from the beginning. This finding corroborates our impression of
phases in the PPM where modelers have a good understanding of the current task.

\noindent\textbf{Challenging Situations.} During the creation of process models,
the two modelers were facing situations they perceived to be more challenging.
This is underpinned by the recorded mean fixation durations. When only considering the
fixations on the process model, TOI V and TOI VII are the most challenging for
M1, since the mean fixation duration is increased, pointing toward higher
attention and a deeper processing of information~\cite{Rau+12}. This seems
reasonable when considering the MPD and CEP's replay. M1 detects an error in the
process model in TOI V. In TOI VII, the mean duration of fixations on the process
model is increased by more than 50ms, i.e., an increase of 29.7\% compared to TOI
VI. In fact, M1 is working on arguably the most challenging part of the process
model, i.e., a long back edge to an earlier part of the process
model~\cite{ReMe11}.  For M2, a similar increase in fixation durations in TOI III
can be observed. The modeler interrupts their modeling endeavour for
additional comprehension and removes some elements from the process model.
In TOI VI of M2, the mean fixation duration is also increased. Notably, M2 is
also working on the challenging part of the process model.
Long mean durations of fixations, however, observed on their own do not
necessarily imply challenging situations. For example, when considering TOI I of
M1, the combination of very long mean duration of fixations on the process model
and the absence of interactions with the modeling environment point toward
inactivity~\cite{Wol+11}.

\noindent\textbf{Causes for Comprehension.} Modelers interrupt their modeling
endeavor for comprehension phases. In a MPD the reason for such a
comprehension phases so far remains in the realm of speculation. On the one
hand, modeler might create the internal representation of the task
description~\cite{PZW+11}. On the other hand, they might have a perfect
understanding of the task, but struggle to convert it into the formal process
model~\cite{PZW+11}. We claim that inspecting the ratio of fixations on the task
description can provide valuable insights. For instance, M2 has several
comprehension phases in TOI III, but the ratio of fixation on the task
description is only 8.2\%. Therefore, we conclude that M2 was rather struggling
with the modeling notation. On the contrary, M1 detects an error in his process
model in TOI V. Similar to M2, several comprehension phases can be identified in
the MPD, but the ratio of fixations on the task description points toward a
different problem. 51.5\% of the fixations are on the task description, the
highest percentage of all TOIs. Therefore, we conclude that M1 had a problem
with the task description instead of the modeling notation.
\vspace{-0.3cm}
\subsection{Outlook} 
\vspace{-0.1cm}

Insights presented in the previous section raise the question whether there are
certain situations in the PPM that are perceived to be more challenging by the
majority of process modelers (with a certain level of experience). This might be
an interesting aspect for future work, since a better understanding of factors
influencing the PPM could be helpful for teaching students in the craft of
modeling. Additionally, we should aim for supporting modelers in challenging
phases of the PPM by providing them specialized tool support rather than
supporting them in phases of fast and focused modeling.

In the near future, we are planning a more detailed eye movement analysis. On
the one hand, we work on separating each PPM instance into phases based on
the part of the process model that is currently edited. This allows us to compare eye
movements of several modelers for a specific part of the process model.
Additionally, a more detailed analysis than comparing fixations on the
textual description and fixations on the process model is in demand. One
interesting aspect might be how often modelers look back to previously created parts of the
process model. Reasons for this might be the validation of previously created
parts, but they could also be looking for similarities to the current problem to
facilitate their problem solving.

%% file: relatedWork.tex
\vspace{-0.4cm}
\section{Related Work}\label{sec:relatedwork}
\vspace{-0.2cm}
Our work is essentially related to model quality frameworks, research on the
PPM and usage of eye movement analysis in conceptual modeling.

Regarding model quality frameworks, there are different frameworks and
guidelines available that define quality for process models. Among others, the
SEQUAL framework uses semiotic theory for identifying various aspects of process
model quality~\cite{krogstie06}, the Guidelines of Process Modeling describe
quality considerations for process models \cite{GOPM}, and the Seven Process
Modeling Guidelines define desirable characteristics of a process model
\cite{DBLP:journals/infsof/MendlingRA10}. While each of these frameworks has been
validated empirically, they rather take a static view by focusing on the
resulting process model, but not on the act of modeling itself. Our research
takes another approach by investigating the process followed to create the
process model.

Research on the process of modeling typically focuses on interactions between
different parties. In a classical setting, a system analyst directs a domain
expert through a structured discussion subdivided into the stages elicitation,
modeling, verification, and validation \cite{hoppenbrouwers05er}. The procedure
of developing process models in a team is analyzed in \cite{rittgen07-caise}.
Interpretation tasks and classification tasks are identified on the semantic
level of modeling. These works build on observation of modeling practice and
distill normative procedures for steering the process of modeling toward a good
completion. Our work, in turn, focuses on the formalization of process models,
i.e., the modeler's interactions with the modeling environment when creating the
formal process model.

In the context of conceptual modeling several experiments have been conducted
investigating the comprehension of UML models, e.g.,~\cite{PoGu10} and the
interpretation of data models, e.g.,~\cite{NoCr99} using eye movement analysis.
In business process management a research agenda has been proposed in~\cite{Hog+11} for investigating
user satisfaction. Our research, focuses on the process of translating an
informal textual description into a formal conceptual model instead of
investigating the comprehension of existing models.

%% file: summary.tex
\vspace{-0.4cm}
\section{Summary}\label{sec:summary}
\vspace{-0.3cm}
This paper demonstrates the feasibility of combining eye movement analysis with
existing research on the PPM to triangulate toward a more comprehensive
understanding of the PPM. Modeling sessions were conducted to collect PPM
instances from students while tracking their eye movements. Based on their MPDs
we selected two examples to illustrate the combination of existing analysis
techniques with eye movement analysis. This combination helps to shed light on
parts of this hardly understood process. Preliminary results revealed insights
into the PPM that could not be obtained by using one analysis technique on its own.
For future work we plan more detailed evaluations with a higher number of
participants to perform quantitative analysis on their PPM instances. We believe
that a better understanding regarding the PPM will be beneficial for future
process modeling environments and will support teachers in mentoring their
students on their way to professional process modelers.

\footnotesize
\noindent\textbf{Acknowledgements.} This research was funded by the Austrian
Science Fund (FWF): P23699-N23.
\vspace{-0.4cm}